\documentclass[epj]{svjour}
\usepackage{graphics}
\begin{document}

\newcommand{\vk}{{\vec k}} 
\newcommand{\vK}{{\vec K}}  
\newcommand{\vb}{{\vec b}}  
\newcommand{\vp}{{\vec p}}  
\newcommand{\vq}{{\vec q}}  
\newcommand{\vQ}{{\vec Q}} 
\newcommand{\vx}{{\vec x}} 
\newcommand{\vh}{{\hat{v}}} 
\newcommand{\cO}{{\cal O}}
\newcommand{\be}{\begin{equation}} 
\newcommand{\ee}{\end{equation}}  
\newcommand{\half}{{\textstyle\frac{1}{2}}}  
\newcommand{\gton}{\mathrel{\lower.9ex \hbox{$\stackrel{\displaystyle 
>}{\sim}$}}}  
\newcommand{\lton}{\mathrel{\lower.9ex \hbox{$\stackrel{\displaystyle 
<}{\sim}$}}}  
\newcommand{\ben}{\begin{enumerate}}  
\newcommand{\een}{\end{enumerate}} 
\newcommand{\bit}{\begin{itemize}}  
\newcommand{\eit}{\end{itemize}} 
\newcommand{\bc}{\begin{center}}  
\newcommand{\ec}{\end{center}} 
\newcommand{\bea}{\begin{eqnarray}}  
\newcommand{\eea}{\end{eqnarray}} 

\title{Heavy quarks at RHIC from parton transport theory}
\author{Denes Molnar \inst{1}\inst{2}
}                     
%
%
\institute{Physics Department, Purdue University, 525 Northwestern Ave, 
West Lafayette, IN 47907, USA \and RIKEN BNL Research Center, Upton, 
NY 11973-5000, USA}
\date{Received: date / Revised version: date}
%
\abstract{
There are several indications that an opaque partonic medium is created in 
energetic Au+Au collisions ($\sqrt{s_{NN}} \sim 100$ GeV/nucleon) at the 
Relativistic Heavy Ion Collider (RHIC). At the extreme densities of 
$\sim 10-100$ 
times normal nuclear density reached even heavy-flavor hadrons are 
affected significantly. Heavy-quark observables are presented
from the parton transport model MPC, focusing on the nuclear 
suppression pattern, azimuthal anisotropy ("elliptic flow"), and 
azimuthal correlations. Comparison with $Au+Au$ data at top RHIC energy 
$\sqrt{s_{NN}} = 200$ GeV indicates significant
 heavy quark rescattering, corresponding roughly 
five times higher opacities than estimates based on
leading-order perturbative QCD. We propose measurements of 
charm-anticharm, e.g., $D$-meson azimuthal correlations as a sensitive, 
independent probe to corroborate these findings.
%
%
%
} 
\maketitle
\section{Introduction}
\label{intro}

Recent heavy-ion collisions experiments at the Relativistic Heavy Ion 
Collider (RHIC) have generated a lot of excitement. Among the most remarkable
discoveries are the large azimuthal momentum anisotropy 
(``elliptic flow'')~\cite{STARv2,PHENIXv2} and
strong attenuation of particles with high transverse momentum created in the 
collision (``jet quenching'')~\cite{PHENIX_RAA,STAR_RAA}, 
which indicate the formation of 
extremely opaque quark-gluon matter that exhibits highly collective,
near-hydro{-}dynamic behavior\cite{WPs}.
The mechanism of rapid randomization and high degree of equilibration 
in the system is not
yet understood.
For example, it is puzzling that 
dissipative effects (such as viscosity) from nonequilibrim transport
are significant\cite{hytrv2} 
and yet ideal (nondissipative) hydrodynamics 
can describe the data quite well\cite{idealhydro}.
For heavy quarks,   
only partial equilibration is expected because
collective effects are weaker due to the large mass.
Therefore, heavy flavor observables are of great interest as an 
orthogonal set of probes to gain more insight and 
cross-check dynamical scenarios.

There are two main dynamical frameworks to study heavy quarks in heavy-ion 
collisions: parton energy loss 
models~\cite{pQCDv2,heavyqRadDK,heavyqRadDG,heavyqCollDj}
and transport approaches%
\cite{Binv2,v2,nonequil,NantesFP,DerekFP,Bin_charmv2,XG}.
Energy loss models consider multiple parton scattering
in the dense medium in an Eikonal approach 
(i.e., straight-line trajectories),
applicable only for very large heavy quark energies.
The advantage, on the other hand, 
is that coherence effects are taken into account.
For charm and bottom, small-angle gluon radiation and therefore 
radiative energy loss is suppressed relative to light quarks
because of the large quark mass
(``dead-cone'' effect)~\cite{heavyqRadDK,heavyqRadDG}. Surprisingly,
recent data from RHIC\cite{PHENIX_e_RAA,STAR_e_RAA} 
indicate little light-heavy difference in the high-$p_T$ suppression pattern.
Though the puzzle is not resolved yet, 
it is clear now that elastic energy loss, previously neglected,
plays an important role~\cite{heavyqCollCU,heavyqCollDj} 
besides radiative energy loss.

Transport models, on the other hand, do not impose kinematic limitations
but typically include incoherent, elastic interactions only.
They are ideal tools to study equilibration because they 
have a hydrodynamic (local equilibrium) limit.
The dynamics is formulated in terms of (on-shell) 6+1D 
phase space distributions that obey the relativitic Boltzmann transport
equation~\cite{Binv2,v2,nonequil,Bin_charmv2,XG},
and the results are mainly sensitive to the transport 
opacity of the system\cite{v2}.
In case particles undergo a lot of scatterings, the evolution for the
bulk of the system (i.e., particles that come from ``typical'' scattering
events and therefore are affected little
by fluctuations in scattering angles
or the number of scatterings)
can be approximated with the Fokker-Planck 
equation\cite{NantesFP,DerekFP}.

In this work we report on heavy flavor (charm and bottom)
observables from covariant transport theory with elastic $2\to 2$ interactions.
The covariant transport solutions were obtained 
using the Molnar's Parton Cascade (MPC) 
algorithm\cite{nonequil,MPC}.
Extending earlier results for charm quark elliptic flow ($v_2$) ~\cite{HQ2004},
we include bottom quarks and also study
heavy nuclear suppression ($R_{AA}$) 
and charm-anticharm azimuthal correlations.
The results are compared to RHIC data and also other transport calculations
in the literature based on the (noncovariant) AMPT transport 
model\cite{Bin_charmv2} or the Fokker-Planck limit\cite{DerekFP}.

\section{Covariant transport theory}
\label{tr}

We consider here  the Lorentz-covariant parton transport theory in
Refs.~\cite{Binv2,v2,ZPC,nonequil,hbt},
in which the on-shell parton phase space densities
$\{ f_i(x,\vp)\}$
evolve with elastic $2\to 2$ rates as
\bea
p_1^\mu \partial_\mu f_{1,i} &=& 
\frac{1}{16\pi^2}\sum\limits_{j} 
\int\limits_2\!\!\!\!
\int\limits_3\!\!\!\!
\int\limits_4\!\!
\left(
f_{3,i} f_{4, j} - f_{1,i} f_{2,j}
\right)
\left|\bar{\cal M}_{12\to 34}^{ij\to ij}\right|^2 
\nonumber \\
&& \qquad\qquad\qquad\qquad \times
\delta^4(p_1+p_2-p_3-p_4)
\nonumber \\
&& + \ S_i(x, \vp_1) \ .
\label{Eq:Boltzmann_22}
\eea
Here $|\bar{\cal M}|^2$ is the polarization averaged scattering matrix 
element squared,
the integrals are shorthands
for $\int\limits_a \equiv \int d^3 p_a / (2 E_a)$,
$g_i$ is the number of internal degrees of freedom for species $i$,
while $f_{a,i} \equiv f_i(x, \vp_a)$.
The source functions $\{S_i(x,\vp)\}$ specify the initial conditions.

Though, in principle, (\ref{Eq:Boltzmann_22}) could be generalized for bosons
and fermions,
or inelastic $3\leftrightarrow 2$ processes~\cite{inelv2,XG},
no practical algorithm yet exists (for opacities at RHIC)
to handle the new nonlinearities such extensions introduce.
We therefore limit our study to quadratic dependence of the collision
integral on $f$.

We apply (\ref{Eq:Boltzmann_22}) to a system of massless gluons and 
light quarks/antiquarks ($q = u,d,s,\bar u, \bar d, \bar s$), and
charm and bottom quarks/antiquarks with mass $m_c = 1.5$ GeV, $m_b = 4.75$ GeV.
All  elastic $2 \to 2$ QCD processes were taken into account: 
$gg\to gg$, 
$gQ \to gQ$, $QQ \to QQ$, and $QQ' \to QQ'$.
Inelastic $2\to 2$ processes, such as $gg\leftrightarrow Q\bar Q$, 
are straightforward 
to include\cite{HQ2004} but were ignored here for faster simulations.

The transport solutions were obtained via Molnar's Parton Cascade 
algorithm~\cite{nonequil,MPC} (MPC),
which employs the parton subdivision technique~\cite{subdivision} 
to maintain Lorentz covariance and causality. Acausal artifacts in
the naive cascade approach (that uses no subdivision) 
are known to affect basic observables such as spectra, elliptic flow,
and freezeout distributions in spacetime\cite{v2,hbt}.

As in Refs.~\cite{Binv2,v2},
only the most divergent parts of the matrix elements 
were considered, regulated using a Debye mass of $\mu_D = 0.7$ GeV.
For perturbative QCD processes at leading-order,
{\em including} scatterings of heavy quarks with gluons and light 
quarks\cite{Combridge},
we thus have
\bea
\frac{d\sigma_{gg\to gg}}{dt} &\approx& 
\frac{9}{4} \frac{d\sigma_{gQ\to gQ}}{dt}
\approx \left(\frac{9}{4}\right)^2 \frac{d\sigma_{QQ'\to QQ'}}{dt}
\nonumber \\
&=& \frac{9\pi \alpha_s^2}{2(t-\mu_D^2)^2} \left(1+\frac{\mu_D^2}{s}\right)
\label{xsec}
\eea
The last expression was obtained assuming a constant
total cross section for $gg\to gg$
(i.e., the weak logarithmic energy dependence was neglected).

In order to reproduce the observed 
elliptic flow for the light parton background at RHIC, 
scattering cross sections between light partons were scaled by a common factor
to obtain $\sigma_{gg\to gg} = 45$ mb~\cite{v2}, 
about fifteen times the elastic $2\to 2$ perturbative QCD estimate.
This value then fixes the total cross sections for all light-parton 
channels. On the other hand, in the spirit of a recent study based on 
Fokker-Planck dynamics\cite{DerekFP},
the enhancement of heavy-quark cross sections was considered 
to be a free parameter.
The motivation for this is that these phenomenological
factors (in part) attempt to incorporate 
the effect of radiative processes, 
which are more important for light partons 
than for the more slowly moving heavy quarks.

The parton initial conditions for
$Au+Au$ at $\sqrt{s_{NN}}=200A$ GeV at RHIC
were similar to those in~\cite{HQ2004},
except that both initial charm and bottom production was, of course, included.
For light partons,
leading order pQCD minijet three-momentum distributions 
were used (with a $K$-factor of 2, GRV98LO PDFs, and $Q^2{=}p_T^2$, while
$Q^2 = \hat s$ for charm).
The low-$p_T$ divergence in the jet cross sections 
was regulated via a smooth extrapolation
below $p_\perp < 2$ GeV 
to yield a total parton $dN(b{=}0)/dy=1000$ at midrapidity.
This choice is motivated by the observed $dN_{ch}/dy \sim 700$ and 
the idea of local parton-hadron duality\cite{Eskola:2000fc}.
More novel hadronization mechanisms, such as parton coalescence,
would imply quite different initial conditions\cite{coalv2}.
Heavy quark momentum distributions were taken from the fixed-order 
plus next-to-leading-log (FONLL) calculation
in~\cite{heavyqRamona},
except for the charm-anticharm correlation results in Sec.~\ref{ccbarcorr}
for which correlated $c-\bar c$ distributions were obtained using the 
PYTHIA event generator\cite{PYTHIA}.
The transverse density distribution was proportional
to the binary collision
distribution for two Woods-Saxon distributions,
therefore $dN^{parton}(b{=}8\ {\rm fm})/dy \approx 250$.
Perfect $\eta=y$ correlation was assumed.

Because heavy quarks are very rare, scatterings between heavy quarks 
and also 
the feedback of heavy quarks on the light-parton background were neglected.
The transport equations (\ref{Eq:Boltzmann_22}) then become 
{\em linear} in the heavy quark phase space 
distributions,
allowing for {\em weighted} test particle sampling 
$f(\vx,\vp,t) = \sum\limits_{i=1}^{N_{test}} w_i \,\delta^3(\vx-\vx_i(t)) 
\,\delta^3(\vp-\vp_i(t))$.
The advantage is that sparse, high-$p_T$ phase-space regions can be sampled 
better (the test particle density can be increased anywhere in phase space,
provided the weight is reduced in inverse proportion).

\section{Results for heavy flavor}
\label{results}

This section contains heavy flavor 
results from the transport model MPC\cite{MPC},
for conditions expected at RHIC. 
The results below are 
labelled by the heavy-quark - gluon scattering cross section
$\sigma$, for which a wide range was explored,
from the leading order perturbative QCD estimate of 
$\sigma \sim 1.3$ mb up to a 15 times enhanced value $\sigma = 20$ mb.

\subsection{Nuclear suppression of charm and bottom}

A common observable to characterize parton energy loss is 
the nuclear suppression factor
$$
R_{AA}(p_T) \equiv
\frac{measured\ yield\ in\ A+A}{expectation\ for\ indep.\ N+N scatterings}
$$
which compares the yields 
to the hypothetical case of independent nucleon-nucleon scatterings.
In this study, the only nuclear affect considered is partonic re-scattering,
therefore the $N+N$ baseline is given by the initial momentum distributions.

\begin{figure*}
\resizebox{0.39\textwidth}{!}{%
  \includegraphics{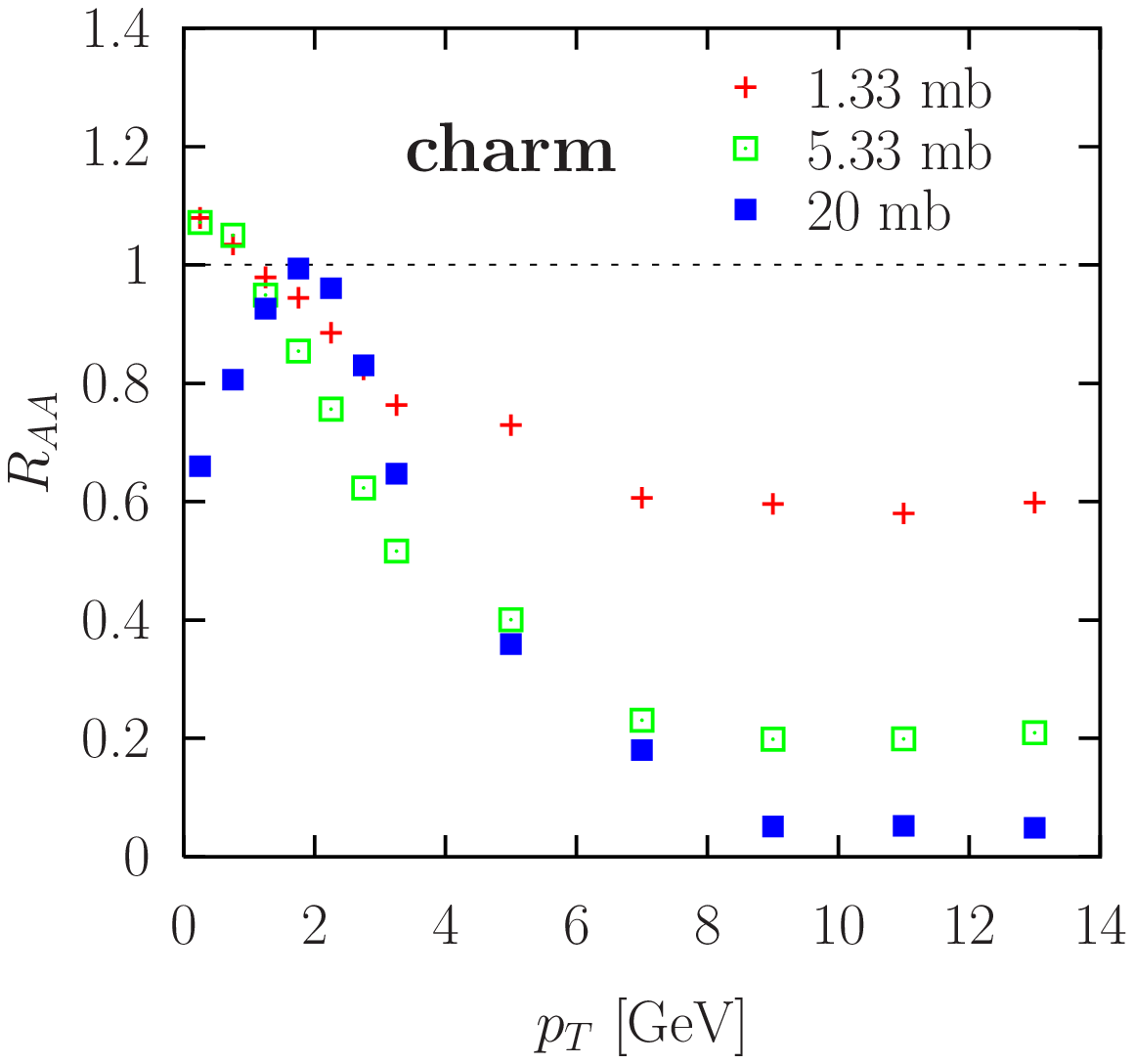}
}
\resizebox{0.39\textwidth}{!}{%
  \includegraphics{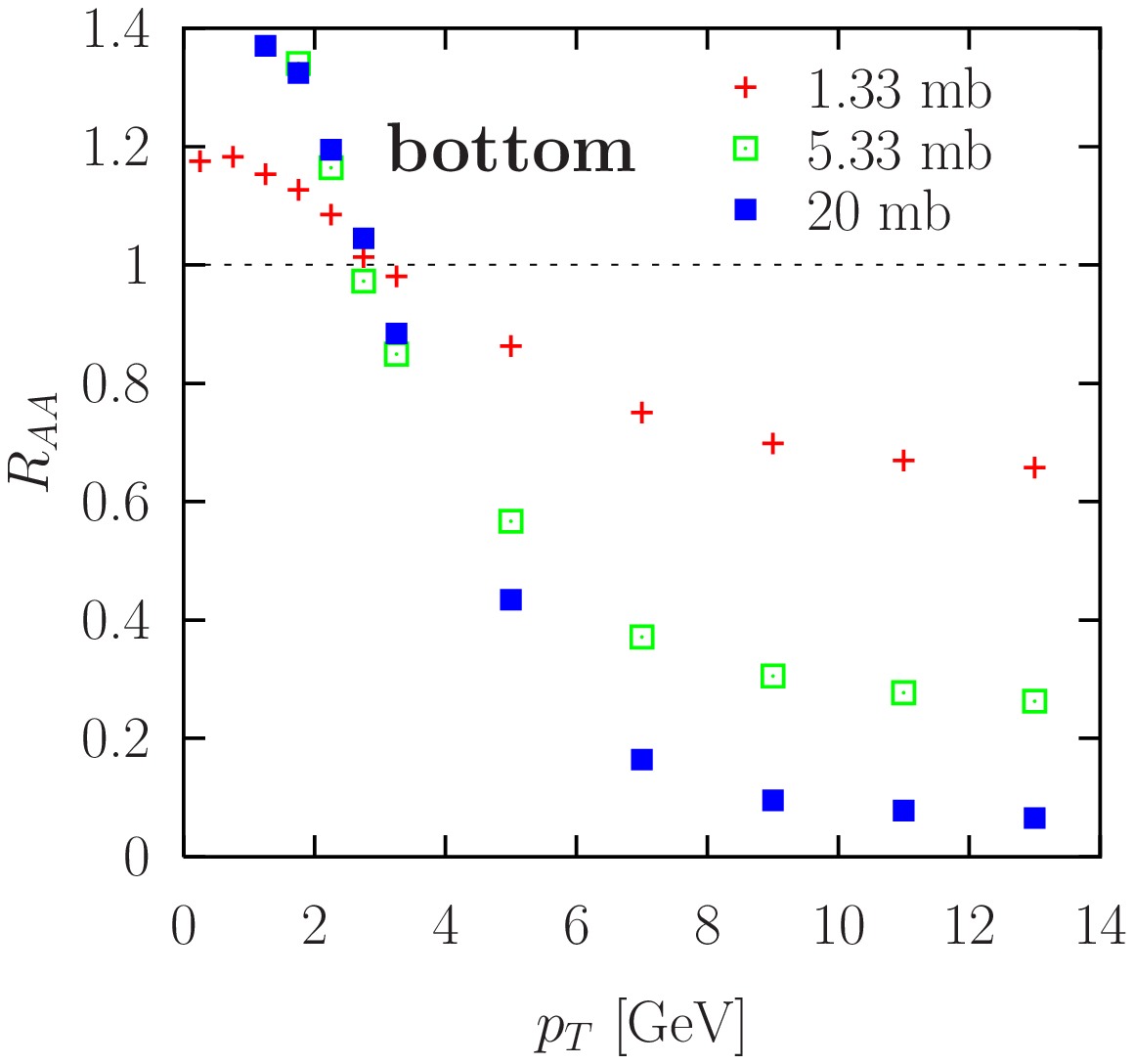}
}
\caption{Nuclear suppression factor $R_{AA}$ for charm (left) and bottom quarks
(right)  as a function of $p_T$ in $Au+Au$ 
at $\sqrt{s_{NN}} = 200$ GeV with $b=8$ fm, 
calculated using the covariant transport model
MPC\cite{MPC} with
heavy-quark - gluon scattering cross sections $\sigma = 1.33$ (pluses), 
$5.33$ (open squares), and $20$ mb (filled squares).
}
\label{fig:1}      
\end{figure*}

Figure \ref{fig:1} shows charm and bottom $R_{AA}$ 
at midrapidity as a function of $p_T$ 
from covariant transport for $Au+Au$
at $\sqrt{s_{NN}}=200$ GeV, with impact parameter $b=8$ fm. 
At high $p_T > 5$~GeV, 
heavy-quark yields are suppressed because of
elastic $2\to 2$ energy loss,
and the suppression becomes stronger with increasing heavy-quark scattering 
cross section $\sigma$. At the same $\sigma$, bottom is less suppressed
than charm, due to the larger bottom mass.
Remarkably, already the
perturbative $\sigma = 1.3$ mb generates a significant suppression 
$R_{AA} \approx 0.6-0.7$. 

At low $p_T$, $R_{AA}$ grows with decreasing $p_T$ 
as a natural consequence of energy loss. 
For the largest $\sigma = 20$ mb, the charm $R_{AA}$ develops a peak
near $p_T \approx 2$ GeV, which is a clear sign of collective flow 
(final spectrum has a ``shoulder-arm'' shape due to radial boost). 
In addition, for such large
cross sections, the midrapidity ($|y| < 1$) 
charm yield is significantly reduced by diffusion in rapidity, 
which is the reason why $R_{AA}$ is below one at all $p_T$.

The charm suppression results are qualitatively similar to 
those from the Fokker-Planck approach in~\cite{DerekFP}.
The main difference is that charm $R_{AA}$ from the 
Fokker-Planck drops much faster as 
$p_T$ increases and does not show any sign of charm diffusion in rapidity.
This is likely because a large {\em final} $p_T$ biases towards fewer
scatterings, moreover,
atypical (``lucky'') scatterings contribute significantly to 
the high $p_T$ yield
\cite{push}. These effects reduce the validity of the 
Fokker-Planck approach at high $p_T$.

\subsection{Elliptic flow of charm and bottom}

\begin{figure*}
\resizebox{0.39\textwidth}{!}{%
  \includegraphics{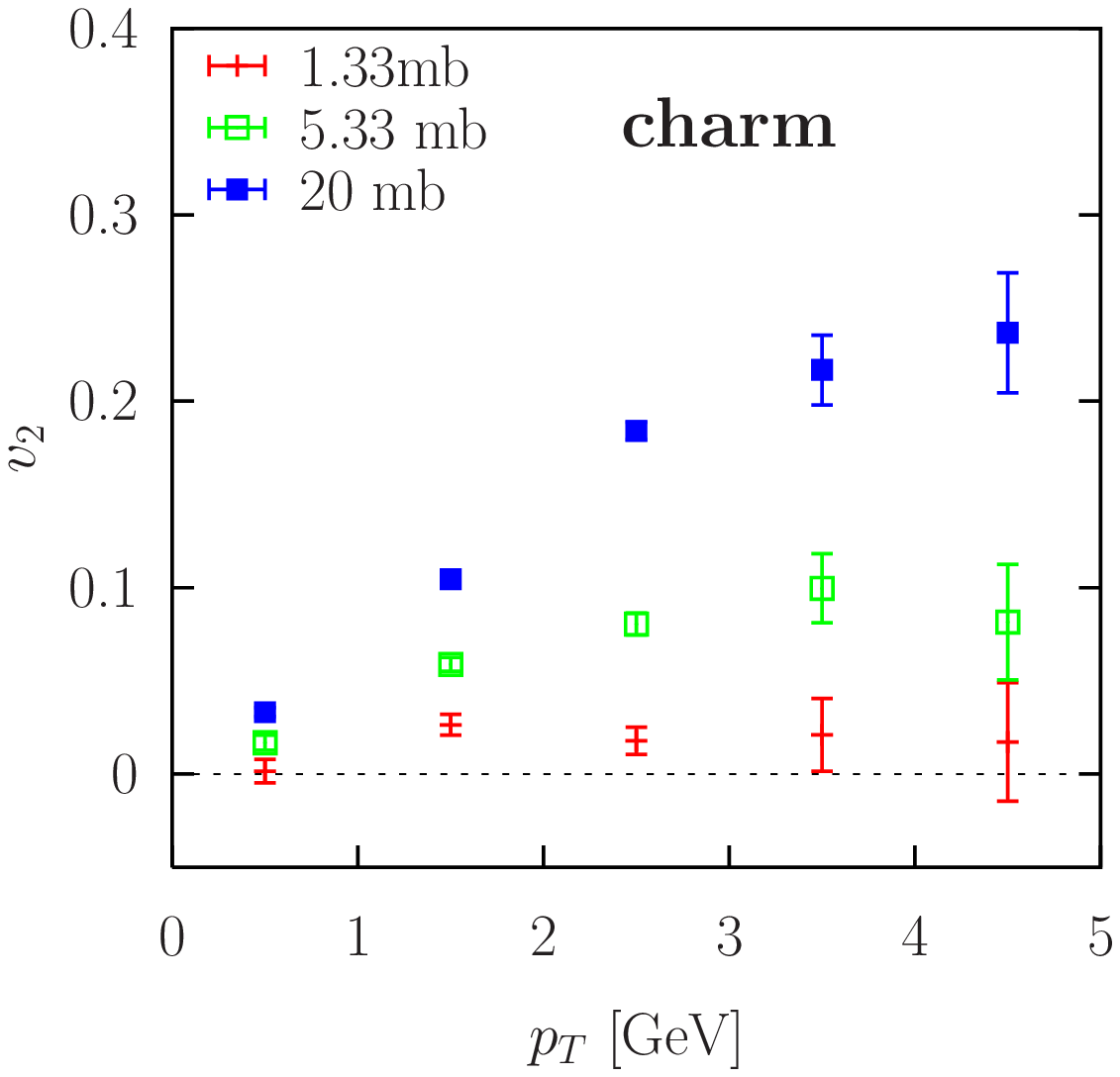}
}
\resizebox{0.39\textwidth}{!}{%
  \includegraphics{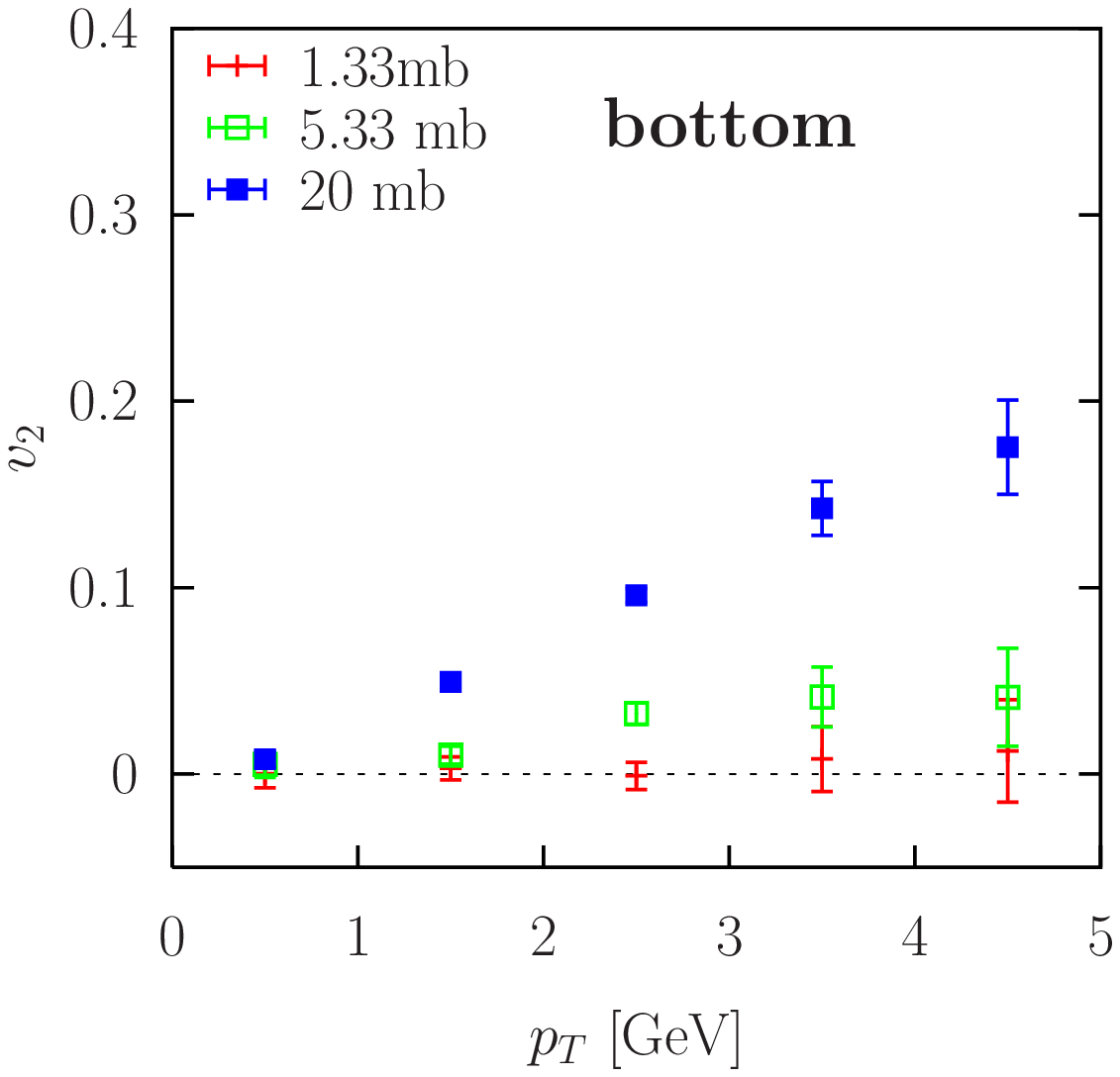}
}
\caption{Differential elliptic flow $v_2$ for charm (left) and bottom quarks
(right) as a 
function of $p_T$ in $Au+Au$ at $\sqrt{s_{NN}} = 200$ GeV with $b=8$ fm,
calculated using the covariant transport model MPC\cite{MPC} with
heavy-quark - gluon scattering cross sections $\sigma = 1.33$ (pluses), 
$5.33$ (open squares), and $20$ mb (filled squares).
}
\label{fig:2}      
\end{figure*}

In noncentral $A+A$ reactions,
an independent measure of 
energy loss and deflections in multiple scatterings is differential 
``elliptic flow'', $v_2(p_T) \equiv \langle \cos (2\phi) \rangle_{p_T}$,
the second Fourier moment of the azimuthal distribution relative
to the reaction plane at a given $p_T$.
Figure~\ref{fig:2} shows charm and bottom $v_2(p_T)$ results at midrapidity 
for $Au+Au$ at
$\sqrt{s_{NN}} = 200$ GeV with $b=8$ fm from covariant transport. 
In the
$0< p_T < 5$ GeV window studied, 
the results are consistent with a monotonic increase with $p_T$ 
for both charm and bottom. For charm, the increase slows down above 
$p_T > 4$ GeV, indicating a turn-over at perhaps $p_T \sim 5-8$ GeV.
At the same $p_T$ and $\sigma$, bottom $v_2$ is below charm $v_2$, as generally
expected from a mass hierarchy\cite{v2cvsh} observed, 
in a lower mass region $m \lton 1.5$ GeV, 
by earlier transport\cite{HQ2004,Bin_charmv2} and ideal hydrodynamic 
calculations\cite{idealhydro}.
For the perturbative estimate $\sigma\sim 1.3$ mb, charm and bottom elliptic 
flow are very small, at most a few percent. Sizeable heavy-quark elliptic flow
$v_2 \sim 0.1$ at moderate $p_T \sim 2-3$~GeV requires 5-10 times enhanced
cross sections.

The charm elliptic flow results agree well with earlier 
results from covariant transport~\cite{HQ2004}, 
and also agree within a factor of 2 with 
results from
the AMPT transport model~\cite{Bin_charmv2}.
The latter calculation considered {\em quark-quark} scattering 
with $3$ mb and $10$ mb cross sections (no gluons) and $2-3$ times higher
parton densities (constituent quarks from the ``string melting'' scenario),
which is roughly 
equivalent to the opacities for $\sigma \sim 6-8$ and $20-25$ mb in our case.
Charm $v_2$ from AMPT tends to be lower
and also saturates earlier, around $p_T \sim 2$ GeV, whereas our results
continue to grow until $p_T \sim 3-4$ GeV.
It would be important to investigate whether
the discrepancy is due to differences in initial conditions, or
the lack of covariance in the AMPT algorithm that has no parton 
subdivision.

The results compare qualitatively well to 
those from the Fokker-Planck approach in~\cite{DerekFP}.
The main difference is that the Fokker-Planck $v_2$ has a higher slope
at low $p_T$ (i.e., much weaker ``mass effect'' for charm) 
and thus saturates earlier at high $p_T$.
In addition, elliptic flow from the transport does not exhibit a
peak (sharp ``rise'' and ``drop'') at moderate $p_T$, 
even for the largest cross section studied here.

\subsection{Suppression and elliptic flow of decay electrons}

Unfortunately, charm and bottom hadrons are difficult to reconstruct 
experimentally. Though various upgrades are under-way to improve detection
capabilities,
the compromise at present is to look at ``non-photonic'' electrons, i.e.,
electrons (and positrons) coming, predominantly, 
from charm and bottom decays.

\begin{figure*}
\resizebox{0.39\textwidth}{!}{%
  \includegraphics{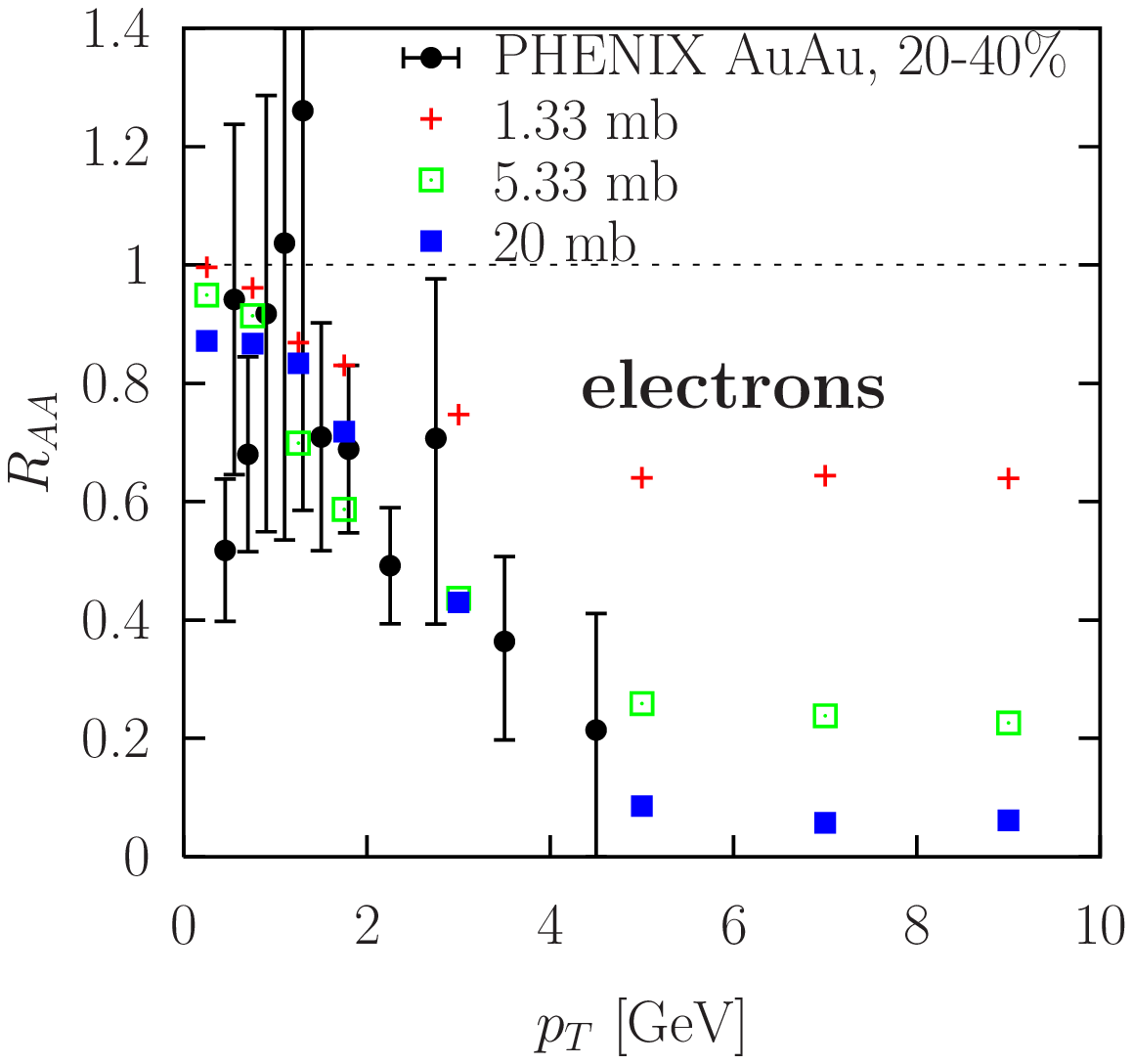}
}
\resizebox{0.39\textwidth}{!}{%
  \includegraphics{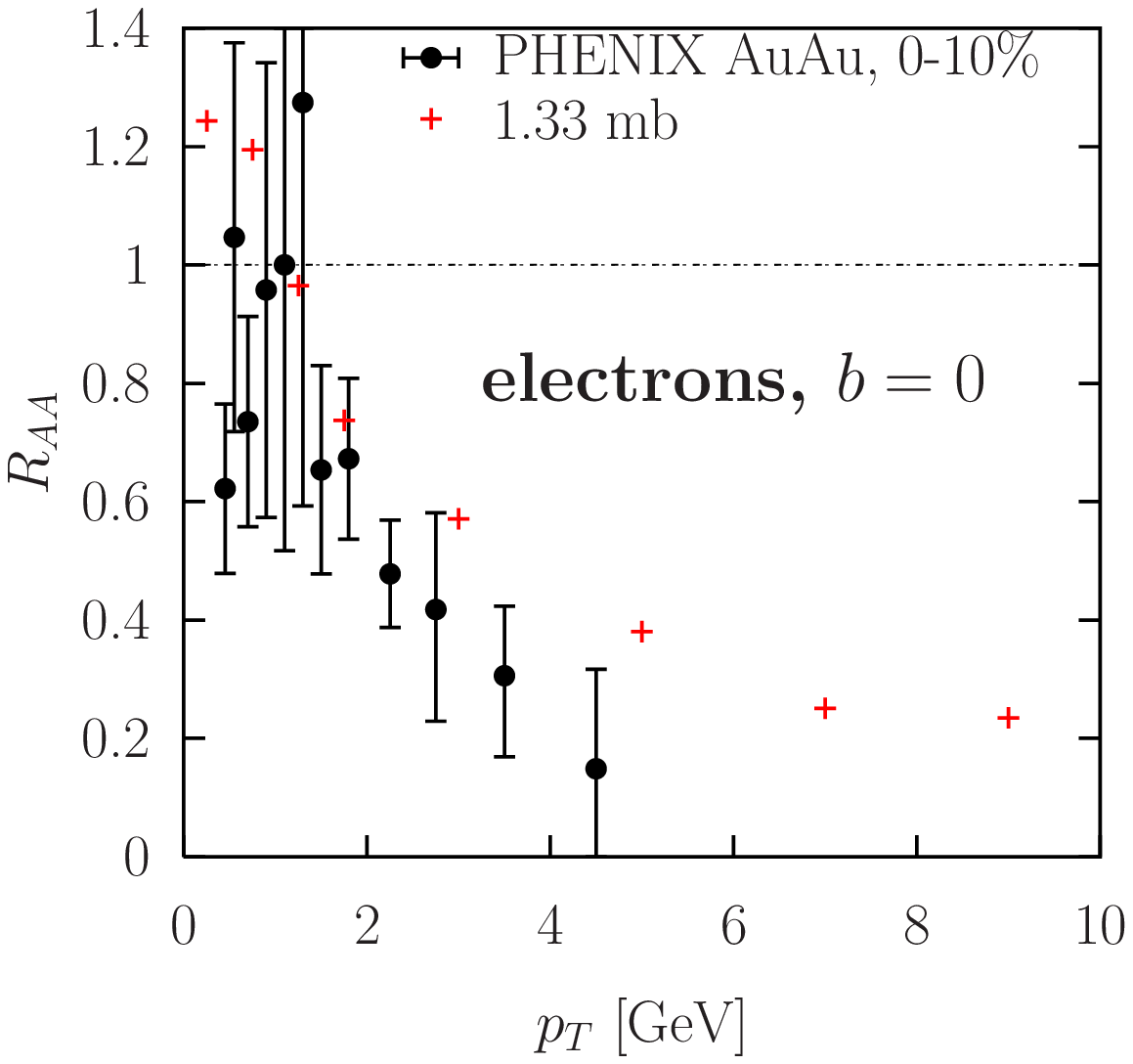}
}
\caption{Left plot shows nuclear suppression factor 
$R_{AA}$ for decay electrons from charm and bottom as a function of $p_T$ in 
$Au+Au$ at $\sqrt{s_{NN}} = 200$ GeV with $b=8$ fm, 
calculated using the covariant transport model
MPC\cite{MPC} with
heavy-quark - gluon scattering cross sections $\sigma = 1.33$ (pluses), 
$5.33$ (open squares), and $20$ mb (filled squares).
The right plot shows the same but for $b=0$ fm and $\sigma = 1.33$ mb only.
Data on ``non-photonic electrons'' 
from PHENIX\cite{PHENIX_e_RAA} (filled circles) with {\em statistical} errors
are also shown.
}
\label{fig:3}      
\end{figure*}

\begin{figure}
\resizebox{0.39\textwidth}{!}{%
  \includegraphics{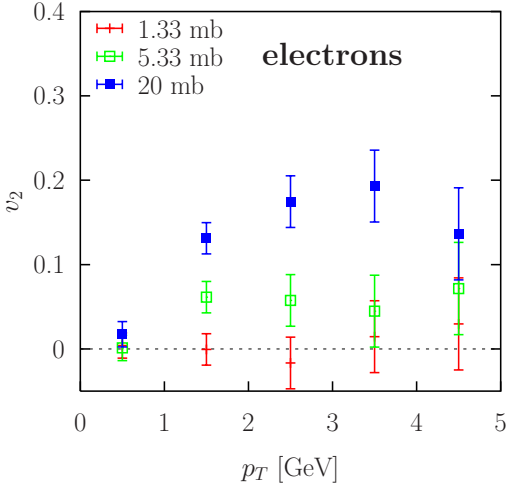}
}
\caption{Differential elliptic flow $v_2$ as a function of $p_T$ 
for decay electrons from charm and bottom in 
$Au+Au$ at $\sqrt{s_{NN}} = 200$ GeV with $b=8$ fm,
calculated using the covariant transport model
MPC\cite{MPC} 
with
heavy-quark - gluon scattering cross sections $\sigma = 1.33$ (pluses), 
$5.33$ (open squares), and $20$ mb (filled squares).}
\label{fig:4}       
\end{figure}

Figures~\ref{fig:3} and \ref{fig:4} show decay electron results from the 
transport for $Au+Au$
at $\sqrt{s_{NN}}=200$ GeV. 
The electron (and positron) spectra were calculated via fragmenting
the heavy quarks into $D$ and $B$ mesons, which were then decayed using 
the PYTHIA event generator\cite{PYTHIA}. 
Data from $d+Au$ collisions at RHIC indicate a very hard 
heavy-quark fragmentation, dominated by momentum fractions 
$z\approx 1$\cite{charmFrag}.
For simplicity, we therefore take fragmentation functions 
$F_{c\to D}(z) = \delta(1-z) = F_{b\to B}(z)$,
and consider only $D^{\pm}$, $D^0$, $\bar D^0$ and the 
corresponding $B$ meson states.

At high $p_T$, electron suppression is very similar in magnitude to that
of heavy quarks, as can be seen in Fig.~\ref{fig:3}. The calculations for
both $b=0$ and 8~fm (about 30\% centrality) indicate insufficient suppression
for perturbative QCD rates. Though experimental uncertainties 
are large, one may speculate that a factor of $\sim 5$ or higher enhancement 
of heavy quark rates is needed for better agreement.
At low $p_T$, much of the structure seen in the heavy quark $R_{AA}$ 
(Fig.~\ref{fig:1}) gets washed out due to the decay kinematics.

Figure~\ref{fig:4} shows the transport results for the elliptic flow
of electrons (and positrons) from charm and bottom decays at midrapidity.
Overall, the electron $v_2(p_T)$ is
very similar to that of charm quarks, except shifted 
to somewhat lower $p_T$ values (as expected from decays).
This corroborates the findings in~\cite{Texascharm} that
only considered electrons from charm decays.
Compared to leading-order perturbative heavy-quark cross sections
that give only a few percent elliptic flow,
a significant $v_2 \sim 5-10$\%
from the transport
requires a $4-8$ times increase in heavy-quark scattering rates
to $\sigma \sim 5-10$ mb.
Based on the electron $R_{AA}$ data, which suggest $\sigma \gton 5$ mb,
one expects an electron $v_2 \gton 5\%$ at moderate $p_T \sim 1.5-5$ GeV.
Preliminary data by STAR\cite{STAR_e_v2} and PHENIX\cite{PHENIX_e_v2} are
compatible with an electron elliptic flow of this magnitude 
but experimental uncertainties unfortunately
prohibit an accurate cross-check.

For the largest $\sigma = 20$ mb from the transport,
at the highest $p_T \approx 4-5$ GeV
the results show a decrease in electron elliptic flow. 
This is because bottom decay
contributions to the overall electron yield
start to become significant (eventually take over at higher $p_T$),
and bottom has a weaker elliptic flow (cf. Fig.~\ref{fig:2}).

\subsection{Charm-anticharm azimuthal correlations}
\label{ccbarcorr}

\begin{figure}
\resizebox{0.39\textwidth}{!}{%
  \includegraphics{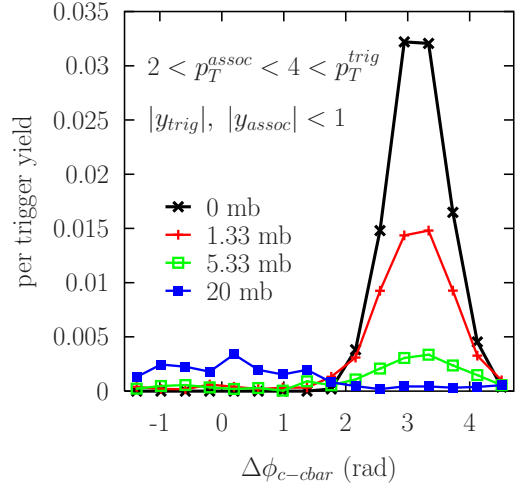}
}
\caption{Charm-anticharm azimuthal correlation
in $Au+Au$ at $\sqrt{s_{NN}} = 200$ GeV with $b=8$,
calculated from the covariant transport model
MPC\cite{MPC} 
with heavy-quark 
- gluon scattering cross sections $\sigma = 1.33$ (pluses), $5.33$
(open squares), and $20$ mb (filled squares).
Charm quarks/anti-quarks  with $p_T > 4$ GeV (triggers)
were correlated with (associated) charm anti-quarks/quarks
with $2 < p_T < 4$ GeV. The correlation without charm rescatterings is also
shown (crosses).
}
\label{fig:5}       
\end{figure}

Rescatterings not only influence the suppression factor and elliptic flow 
but also the azimuthal
correlations between two heavy quarks.
Figure~\ref{fig:5} shows the charm-anticharm 
correlation pattern expected for $Au+Au$
at $\sqrt{s_{NN}} =200$ GeV with $b=8$ fm from covariant 
transport. In the calculation of this observable, the (correlated) 
initial charm distributions were taken from PYTHIA\cite{PYTHIA}.
PYTHIA predicts a strong away-side peak 
in $N+N$ collisions (i.e., for $\sigma = 0$).
However, 
the transport results show a correlation strength that  
is very
sensitive to heavy-quark rescattering in heavy-ion collisions.
The away-side peak is 
already reduced by about half for the small perturbative value
$\sigma \approx 1.3$ mb, and as the cross section is increased further, 
the peak
rapidly weakens and broadens. Eventually, for very large $\sigma$, 
the correlation changes character and a very broad 
near-side peak appears. 
Measurements of charm-anticharm, such as, $D$-meson 
azimuthal correlations
can therefore provide an independent
way to determine the effective heavy quark scattering rates 
in the dense nuclear medium 
formed in heavy-ion collision. 

\section{Conclusions}
\label{concl}

In this work we present heavy-flavor observables in $Au+Au$
at RHIC (mainly for impact parameter $b = 8$ fm) 
from covariant parton transport theory. The heavy-quark phase space
evolution was studied in an order of magnitude
more opaque light parton (quark and gluon) 
system than a perturbative parton gas
with leading-order $2\to 2$ interactions\cite{v2}. The calculation was
driven by the cross section $\sigma$ of heavy-quark interactions with gluons. 
The transport solutions
were obtained using the covariant MPC algorithm\cite{nonequil,MPC}. 

We find significant charm and bottom suppression with $R_{AA} \sim 0.5-0.65$ 
at high $p_T > 4$ GeV already for leading-order
heavy-quark matrix elements, and a decreasing $R_{AA}$ as the cross section 
increases (Fig.~\ref{fig:1}). 
Electrons from heavy-quark decays 
show a suppression pattern very similar in magnitude
to that of the heavy quarks (Fig.~\ref{fig:3}). Consistency with data 
from RHIC\cite{PHENIX_e_RAA,STAR_e_RAA} on ``non-photonic'' electrons requires
a roughly five-fold increase in heavy-quark opacities relative to perturbative
$2\to 2$ scattering (more precise data would give better constraints). 

Charm and bottom elliptic flow $v_2(p_T)$ 
from covariant transport is at most a few percent
for the perturbative estimate of $\sigma\sim 1.3$ mb. Significant heavy quark 
elliptic flow $v_2 \gton 0.1$ at moderate $p_T \sim 2-3$ GeV requires about
five-fold or more enhanced heavy quark rescattering (Fig.~\ref{fig:3}). 
Electron $v_2(p_T)$ is
very similar to that of charm, at least up to $p_T \approx 5$ GeV, where bottom
contributions to the decay electron yield start to become important.
For heavy-quark opacities indicated by the non-photonic 
electron $R_{AA}$ data at RHIC,
the transport predicts $v_2 \sim 5-10$\% (Fig.~\ref{fig:4}).
This is within the large uncertainties of 
current measurements by STAR\cite{STAR_e_v2} and PHENIX\cite{PHENIX_e_v2} - 
more accurate data are highly desirable.

In addition, we propose a unique observable, charm-anticharm azimuthal 
correlations, as an independent, sensitive probe of the degree of 
charm rescatterings in the dense parton medium (Fig.~\ref{fig:5}). 
High $p_T > 4$ GeV 
charm/anti-charm
quarks (triggers) were correlated with moderate $2 < p_T < 4$ GeV 
anticharm/charm quarks, both at midrapidity.
The strong away-side correlation peak in this observable
predicted by PYTHIA for $N+N$ collisions is strongly reduced 
(and also broadened) due to rescatterings, by already a factor of two 
for the small perturbative cross section $\sigma \sim 1.3$ mb.
At very large cross sections, the correlation pattern even changes to a broad
near-side peak.

We emphasize that this study is limited to $2\to 2$ transport. Contributions
from radiative channels are likely important and should be included in the 
future. The results, nevertheless, can serve as a baseline calculation 
of elastic energy loss effects.

Finally, it would be interesting to extend this calculation
with hidden heavy-flavor observables, such as $J/\psi$ suppression,
for which data are also available from RHIC.

\medskip
{\bf Acknowledgments.} We thank RIKEN, Brookhaven National Laboratory and
the US Department of Energy [DE-AC02-98CH10886] for providing facilities
essential for the completion of this work.

%

\end{document}